\documentclass[aps,prl,twocolumn,superscriptaddress,showkeys,10pt]{revtex4-1}

\usepackage[utf8]{inputenc}
\usepackage{graphicx}
\usepackage{epsfig}
\usepackage{amsmath,amsthm}
\usepackage{color}
\usepackage{verbatim}
\usepackage{ulem}

\begin{document}

%\title{\color{black} Direct Observation of Time Reversal Symmetry Breaking in %Magnetic Circular Dichroism in 
%Altermagnetic RuO$_2$ Band Structure by Photoemission Spectroscopy}

\title{\color{black} Observation of time-reversal symmetry breaking in the band structure of altermagnetic  RuO$_2$}

\author{O. Fedchenko}
\affiliation{Institut f\"{u}r Physik, Johannes Gutenberg-Universit\"{a}t Mainz, Staudingerweg 7, D-55099 Mainz, Germany}
\author{J. Min\'ar}
\affiliation{University of West Bohemia, New Technologies Research Center, Plzen 30100, Czech Republic}
\author{A. Akashdeep}
\affiliation{Institut f\"{u}r Physik, Johannes Gutenberg-Universit\"{a}t Mainz, Staudingerweg 7, D-55099 Mainz, Germany}
\author{S.W. D'Souza}
\affiliation{University of West Bohemia, New Technologies Research Center, Plzen 30100, Czech Republic}
\author{D. Vasilyev}
\affiliation{Institut f\"{u}r Physik, Johannes Gutenberg-Universit\"{a}t Mainz, Staudingerweg 7, D-55099 Mainz, Germany}
\author{O. Tkach}
\affiliation{Institut f\"{u}r Physik, Johannes Gutenberg-Universit\"{a}t Mainz, Staudingerweg 7, D-55099 Mainz, Germany}
\affiliation{Sumy State University, Rymski-Korsakov 2, 40007 Sumy, Ukraine}
\author{L. Odenbreit}
\affiliation{Institut f\"{u}r Physik, Johannes Gutenberg-Universit\"{a}t Mainz, Staudingerweg 7, D-55099 Mainz, Germany}
\author{Q.L. Nguyen}
\affiliation{Linac Coherent Light Source, SLAC National Accelerator Laboratory, Menlo Park, CA 94025, USA}
\author{D. Kutnyakhov}
\affiliation{Ruprecht Haensel Laboratory, Deutsches Elektronen-Synchrotron DESY, 22607 Hamburg, Germany}
\author{N. Wind}
\affiliation{Ruprecht Haensel Laboratory, Deutsches Elektronen-Synchrotron DESY, 22607 Hamburg, Germany}
\author{L. Wenthaus}
\affiliation{Ruprecht Haensel Laboratory, Deutsches Elektronen-Synchrotron DESY, 22607 Hamburg, Germany}
\author{M. Scholz}
\affiliation{Ruprecht Haensel Laboratory, Deutsches Elektronen-Synchrotron DESY, 22607 Hamburg, Germany}
\author{K. Rossnagel}
\affiliation{Institut für Experimentelle und Angewandte Physik, Christian-Albrechts-Universität zu Kiel, 24098 Kiel, Germany}
\affiliation{Ruprecht Haensel Laboratory, Deutsches Elektronen-Synchrotron DESY, 22607 Hamburg, Germany}
\author{M. Hoesch}
\affiliation{Ruprecht Haensel Laboratory, Deutsches Elektronen-Synchrotron DESY, 22607 Hamburg, Germany}
\author{M. Aeschlimann}
\affiliation{Universit\"{a}t Kaiserslautern, Department of Physics, 67663 Kaiserslautern, Germany}
\author{B. Stadtm\"{u}ller}
\affiliation{Institut f\"{u}r Physik, Johannes Gutenberg-Universit\"{a}t Mainz, Staudingerweg 7, D-55099 Mainz, Germany}
\author{M. Kl\"{a}ui}
\affiliation{Institut f\"{u}r Physik, Johannes Gutenberg-Universit\"{a}t Mainz, Staudingerweg 7, D-55099 Mainz, Germany}
\author{G. Sch{\"o}nhense}
\affiliation{Institut f\"{u}r Physik, Johannes Gutenberg-Universit\"{a}t Mainz, Staudingerweg 7, D-55099 Mainz, Germany}
\author{G. Jakob}
\affiliation{Institut f\"{u}r Physik, Johannes Gutenberg-Universit\"{a}t Mainz, Staudingerweg 7, 
D-55099 Mainz, Germany}
\author{T. Jungwirth}
\affiliation{Inst. of Physics Academy of Sciences of the Czech Republic, Cukrovarnick\'{a} 10,  Praha 6, Czech Republic}
\affiliation{School of Physics and Astronomy, University of Nottingham, NG7 2RD, Nottingham, United Kingdom}
\author{L. \v{S}mejkal}
\affiliation{Institut f\"{u}r Physik, Johannes Gutenberg-Universit\"{a}t Mainz, Staudingerweg 7, D-55099 Mainz, Germany}
\affiliation{Inst. of Physics Academy of Sciences of the Czech Republic, Cukrovarnick\'{a} 10,  Praha 6, Czech Republic}
\author{J. Sinova}
\affiliation{Institut f\"{u}r Physik, Johannes Gutenberg-Universit\"{a}t Mainz, Staudingerweg 7, D-55099 Mainz, Germany}
\affiliation{Inst. of Physics Academy of Sciences of the Czech Republic, Cukrovarnick\'{a} 10,  Praha 6, Czech Republic}
\author{H. J. Elmers}
\affiliation{Institut f\"{u}r Physik, Johannes Gutenberg-Universit\"{a}t Mainz, Staudingerweg 7, D-55099 Mainz, Germany}
\email{elmers@uni-mainz.de}

\date{\today}

\begin{abstract}
\color{black}
Altermagnets are an emerging third elementary class of magnets. Unlike ferromagnets, their distinct crystal symmetries inhibit magnetization while, unlike  antiferromagnets, they promote strong spin polarization in the band structure. The corresponding unconventional mechanism of time-reversal symmetry breaking  without magnetization in the electronic spectra has been regarded as a primary signature of altermagnetism, but has not been experimentally visualized to date.  We directly observe  strong time-reversal symmetry breaking in the band structure of altermagnetic RuO$_2$ by detecting magnetic circular dichroism in angle-resolved photoemission spectra. Our experimental results, supported by ab initio calculations, establish the microscopic electronic-structure basis for  a  family of novel  phenomena and functionalities in fields ranging from topological matter to spintronics, that are based on the unconventional time-reversal symmetry breaking in altermagnets.  
\end{abstract}

\maketitle

\section{Introduction}
%The recent discovery of altermagnetism  directly provides a viable physical path to merge in one material  class several of the advantageous properties of ferromagnets and antiferromagnets for future spintronic device concepts \cite{Smejkal2021a,Smejkal2022a}. Altermagnets exhibit an unconventional spin-polarized $d/g/i$-wave electronic structure in reciprocal space, originating from local anisotropies in direct space of the opposite-spin subllattices connected by a rotation symmetry. This gives rise to properties unique to altermagnets (e.g., the spin-splitter effect \cite{GonzalezHernandez2021}, the crystal anomalous Hall effect \cite{Smejkal2020}, and chiral magnons in zero magnetization systems \cite{Smejkal2022b}), as well as certain properties of ferromagnets (e.g., spin-polarized currents) and antiferromagnets (e.g., THz spin dynamics and vanishing net magnetization). 

Conventionally, two elementary classes of crystals with collinear magnetic order have been considered - ferromagnetic and antiferromagnetic. The ferromagnetic exchange interaction generates strong magnetization and spin-polarization in electronic bands that break time-reversal ($\cal{T}$) symmetry. Non-dissipative Hall currents, including their topological quantum variants \cite{Nagaosa2010,Tokura2019,Smejkal2021b}, as well as spin-polarized currents, %that 
%enable highly efficient reading and writing of information 
vital in modern ferromagnetic information technologies \cite{Chappert2007,Ralph2008,Bader2010,Bhatti2017}, are all based on the strong $\cal{T}$ symmetry breaking in the electronic structure. However, ferromagnetic and topological-insulating phases are %mutually 
poorly compatible, and the inherent magnetization of ferromagnets limits the capacity and speed of ferromagnetic spintronic devices. In the second conventional class, the antiferromagnetic exchange generates compensated collinear order with no magnetization. The resulting absence in antiferromagnets of %the ferromagnetic-like 
strong $\cal{T}$ symmetry-breaking linear responses akin to ferromagnets has forced the antiferromagnetic spintronic research to exploit comparatively weak phenomena relying on relativistic spin-orbit coupling \cite{Jungwirth2016,Manchon2019}. On the one hand, the weak responses represent a roadblock. On the other hand, the zero magnetization is well compatible with materials ranging from superconductors to insulators, and it enables breakthroughs towards information technologies with ultra-high capacity and speed \cite{Jungwirth2018,Baltz2018,Kimel2019}. 

The above examples illustrate why discoveries of magnetic quantum matter with unconventional characteristics and functionalities remain central to the frontier research in condensed-matter physics, and to the development of ultra-scalable low-power technologies. Recently, a symmetry classification and description, focused within  the hierarchy of  interactions on the strong exchange, has led to the identification of a third elementary class of crystals with a collinear magnetic order, dubbed altermagnets \cite{Smejkal2021a,Smejkal2022a}. As illustrated in Fig. 1, altermagnets have a symmetry-protected compensated antiparallel magnetic order on a crystal that generates an unconventional alternating spin-polarization and $\cal{T}$ symmetry breaking in the band structure without magnetization \cite{Smejkal2020}.  Altermagnets have been thus predicted to combine merits of ferromagnets and antiferromagnets, that were regarded as principally incompatible, and to have merits unparalleled in either of the two conventional magnetic classes \cite{Smejkal2021a,Smejkal2022a}. While unconventional anomalous Hall  and spin-polarized currents have been predicted \cite{Smejkal2021b,Smejkal2022a,Smejkal2020,Samanta2020,Naka2020,Hayami2021,Mazin2021,Betancourt2021,Naka2022,Naka2019,Gonzalez-Hernandez2021,Naka2021,Ma2021,Smejkal2022,Smejkal2021a} and recently observed in experiment \cite{Feng2022,Betancourt2021,Bose2022,Bai2022,Karube2022},
%, such as anomalous Hall effect \cite{Feng2022}, tilted spin-currents \cite{Bose2022} and the spin-splitter torque \cite{Bai2022,Karube2022}, 
so far there has  not been a direct measurement of the underlying $\cal{T}$ symmetry breaking in the altermagnetic band structure. 

\begin{figure*}[t]
\includegraphics[width=0.7\textwidth]{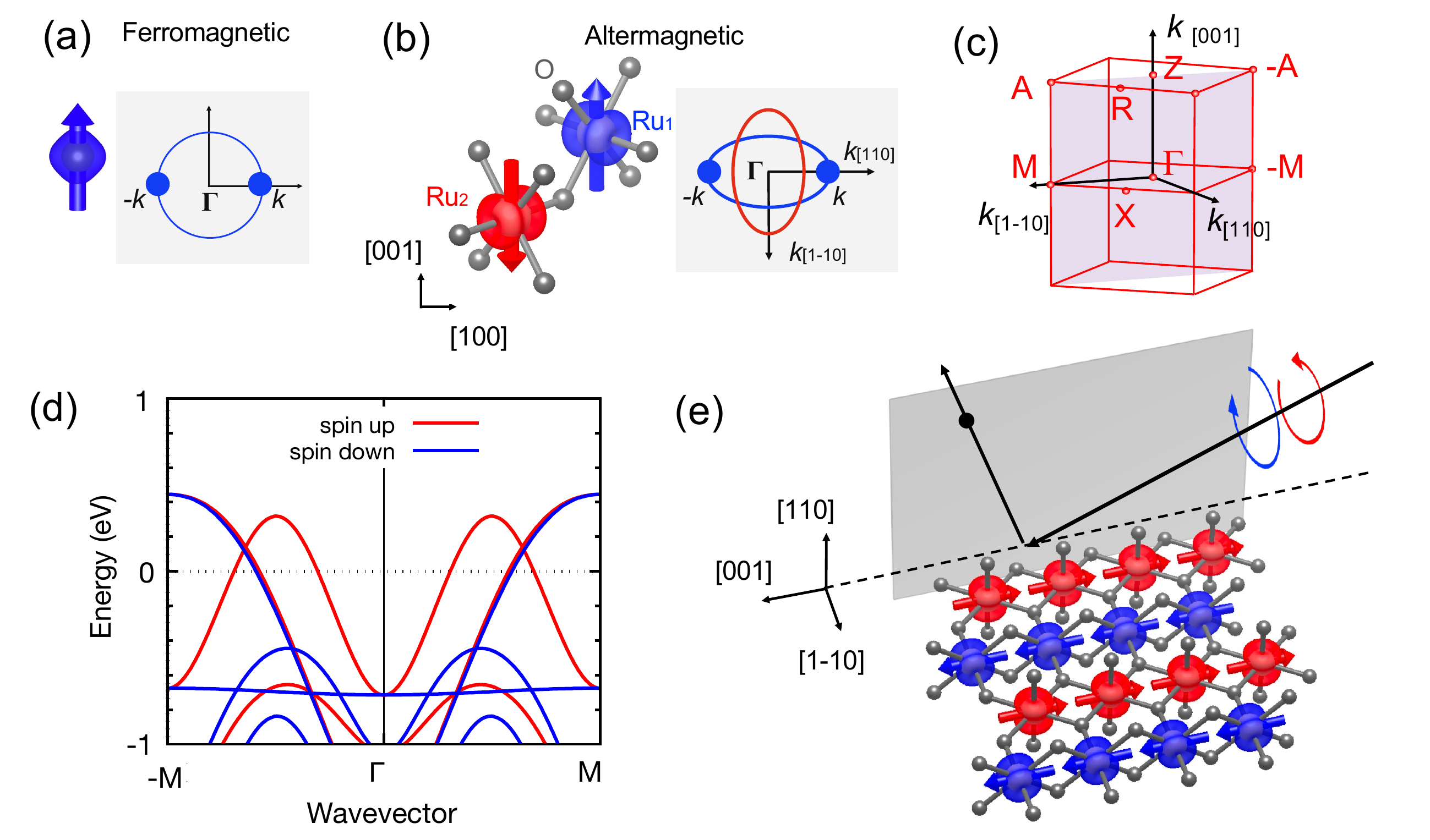} 
\vspace{-0,5 cm}
\caption{\label{Fig0} 
Illustrative comparison of (a) ferromagnetism,  whose magnetization (left) generates conventional spin-polarization and $\cal{T}$-symmetry breaking in the band structure (right), and (b) altermagnetism, whose symmetry-protected compensated antiparallel magnetic order on a crystal (left) generates an unconventional alternating spin-polarization and $\cal{T}$-symmetry breaking in the band structure without magnetization (right) \cite{Smejkal2021a,Smejkal2022a}. Color-coding of bands in the momentum space reflects the spin orientation as depicted by arrows in the real space. $\cal{T}$-symmetry in the band structure is broken since time-reversed (opposite) momenta on the energy isosurfaces have non-time-reversed (same) spin orientations. In (b), the real-space model corresponds to RuO$_2$ with grey spheres representing O-atoms and color-surfaces representing magnetization densities on Ru atoms. (c) Brillium zone indicating the high symmetry points relevant for the spectra shown in the data.  
(d) Ab initio calculation of the band structure of RuO$_2$, showing a strongly broken $\cal{T}$ symmetry. %Red and blue indicate the spin polarization of the bands. 
(e) Sketch of the RuO$_2$ magnetic crystal structure with the surface oriented along the [110] direction, and the experimental set-up of scattered photoelectrons from circularly polarized light. }
\vspace{-0,0 cm}
\end{figure*}
A suitable microscopic tool  is based on magnetic circular dichroism (MCD) which is an optical counterpart of the $\cal{T}$ symmetry breaking anomalous Hall effect. The presence of MCD in altermagnets over the full spectral range up to X-rays has been confirmed by ab initio calculations \cite{Samanta2020,Hariki2023}. MCD combined with angle-resolved photoemission spectroscopy  then allows for the microscopic visualization of the $\cal{T}$ symmetry breaking in the electronic-structure in momentum space. In the past, this technique has been successfully used in the investigation of  ferromagnetic materials~\cite{Schneider1991,Bansmann1992,Stoehr1993,Laan1993,Braun1996,Ebert1997,Henk1996,Yokoyama2008,Hild2009}.

Our experimental study focuses on epitaxial RuO$_2$, 
a workhorse material of the altermagnetic class \cite{Smejkal2021a,Smejkal2022a}. 
The  rutile crystal structure of metallic RuO$_2$ has been shown to have a collinear compensated magnetic order~\cite{Berlijn2017,Zhu2019}, and 
predicted to be an altermagnet ~\cite{Smejkal2021a,Ahn2019} with a strong (order eV) $\cal{T}$ symmetry breaking spin-splitting in momentum space, as shown in~Fig.~\ref{Fig0}(b) and (d).
This rutile crystal family is predicted to show topological properties as well, with prior spin-integrated angle-resolved photoemission studies reporting  two Dirac nodal lines and pronounced topological interface states \cite{Jovic2018}. More recently, %as we have already noted, 
the predictions of strong anomalous Hall and spin currents  in combination with vanishing magnetization \cite{Smejkal2020,Gonzalez-Hernandez2021} have  also  been  experimentally verified~\cite{Feng2022,Bose2022,Bai2022,Karube2022} in this  altermagnet.
% as well as observations of
%strain-induced superconductivity~\cite{Gregory2022,Ruf2021}.
%In addition, the predicted strong spin-splitting in RuO$_2$ of non-relativistic origin, can also lead to giant or tunneling magnetoresistance devices \cite{Shao2021,Smejkal2022}, similar to their ferromagnetic counterparts.

Here we present  direct evidence for a strong $\cal{T}$ symmetry breaking in the band structure of
 epitaxial altermagnetic RuO$_2$ films
 by  detecting MCD in the angular distribution of photoelectrons, both for soft X-ray and ultraviolet photon excitation. We compare the experimental results to the corresponding calculations based on  density functional theory. 
%confirming the existence of a strong spin-splitting in the RuO$_2$ band structure.
%Our results demonstrate that the MCD effect provides a powerful tool for the investigation of altermagnetic band structures, exhibiting the unconventional  spin polarization and $\cal{T}$ symmetry breaking.

%\section{Experimental Methods}
\begin{figure*}
\includegraphics[width=0.8\textwidth]{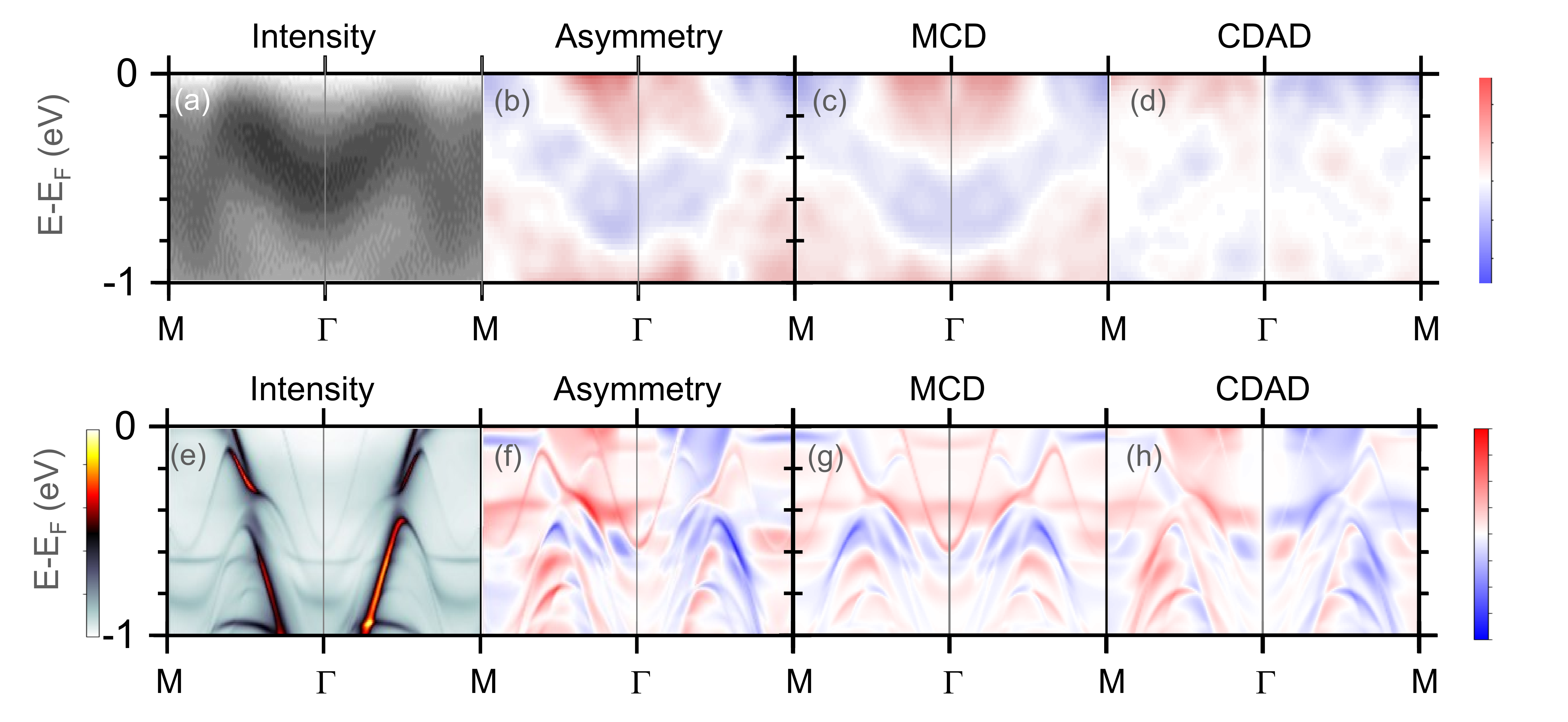}
\vspace{-0,6 cm}
\caption{\label{Fig2JS} 
{(a) Measured intensity map $I(E_B,k_y)$ revealing the energy dispersion of bands along the line 
M-$\Gamma$-M (See Fig. 1 (c)) measured at 70~K. Here dark is high intensity. (b) Measured intensity asymmetry $A(E_B,k_y)$, (c) MCD  
 $A_{\rm MCD}(E_B,0,k_y)$ and (c) CDAD  $A_{\rm CDAD}(E_B,0,k_y)$ extracted from the  asymmetry. 
(e) Calculated spectral density average and (f) its asymmetry for + and - polarized light within the experimental geometry. The theoretical
(g) MCD and (h) CDAD are extracted from the spectral density asymmetry as with the experimental MCD and CDAD. The photon energy used here is 660 eV.}
}
\vspace{-0,6 cm}
\end{figure*}

{\it Experimental Methods} -- We have grown epitaxial RuO$_2$(110) films with a thickness of 34~nm  by pulsed laser deposition on
TiO$_2$(110) substrates that were heated during deposition to 400 $^o$C. The samples show no detectable remanent magnetization, consistent with the earlier magnetometry studies of analogous RuO$_2$/TiO$_2$ thin films \cite{Feng2022}.
%The oxygen pressure was 0.02~mbar and typical growth rates were 1.9~nm/min with the KrF excimer laser running at 10~Hz and 150~mJ pulse energy. Samples 
% were structurally characterized by X-ray diffraction, X-ray reflectometry and in-situ reflective high energy electron diffraction (RHEED). From  X-ray reflectometry we find that 34~nm samples have a surface roughness smaller than 0.3~nm, while the width of the out-of plane scattering peak in 2$\Theta$ indicates a somewhat reduced coherent scattering volume of 22~nm. The existence of the RHEED scattering pattern shows that crystallinity persists up to the surface. The full width at half maximum of the rocking curve is 1.0$^o$ indicating a good alignment of the surface normal. $\Phi$ scans reveal the epitaxial growth and show a two fold symmetry of (200) peaks aligned with that from the substrate. The width of the peaks in $\Phi$ is 2$^o$. From the determined lattice constants we conclude on a compressive strain of -1.4\% along $c$-axis and a tensile strain of 0.4\% along $a$-axis. 
%Samples were transported from the deposition chamber to the photoemission experiment 
%using an ultra-high vacuum suitcase.
For growth and sample characterization details see Supplementary Materials.

For the photoemission measurements,  photoelectrons were excited by  circularly-polarized soft X-rays 
%a He discharge lamp (21.2~eV), by a pulsed laser (6.4~eV, 80~MHz repetition rate, APE), and by soft X-ray radiation
(beamline P04, PETRA III, DESY, Hamburg). For these %soft X-ray 
experiments
we used the 
time-of-flight momentum microscope DRUMSOX installed at the open port I of the beamline P04 with an energy resolution of 60~meV at a sample temperature of 70~K.
In addition, circularly polarized ultraviolet light by a pulsed laser (6.4~eV, 80~MHz repetition rate, APE) was used.
The photoemission experiments with laser excitation have been conducted using a 
time-of-flight momentum microscope (ToFMM, Surface Concept GmbH) with the resolution set to 
40~meV~\cite{Medjanik2017} and at 20~K.

The circular-dichroism photoemission experiments described below have been performed with the incidence angle of the photon beam at 22$^\circ$ with respect to the sample surface and the azimuthal orientation of the sample has been adjusted so that the photon incidence plane coincides with the easy spin axis of RuO$_2$, i.e., the [001] c-axis \cite{Berlijn2017a,Smejkal2020,Feng2020a}.

%Photoemission experiments at 21.2~eV have been performed using the single-hemisphere momentum microscope described in Ref.~\cite{Schoenhense2020} with the energy resolution set to 50~meV and at a sample temperature of 20~K. 
The coordinate system for the photoelectron momentum ($k_x,k_y,k_z$) is set to $k_z$ along  the crystallographic [110] direction, i.e. surface normal, $k_x$ along [001] and $k_y$ along  {$[1\bar{1}0]$} in-plane directions, respectively.  
A sketch of the experimental geometry is shown in Fig.~\ref{Fig0} (e).

%\section
{\it Results with soft X-ray excitation} -- Using soft X-ray excitation,
we can measure the intensity distribution of the direct transitions in four-dimensional energy-momentum space $I(E_B,k_x,k_y,k_z)$, %, which is simultaneously recorded in the time-of-flight momentum microscope. 
%$I(E_B,k_x,k_y,k_z)$ 
which is the spectral density function modulated by matrix elements accounting for the photo-excitation probability for a given initial $k_i$ and final state $k_f$.
As described in detail in the Supplementary Materials,
%Considering the energy conservation and the role of reciprocal
%lattice vectors in order to obey momentum conservation,
for a given photon energy $h\nu$ and binding energy $E_B=E-E_F$,
the final photoelectron states are located on a 
spherical shell with radius (for units \AA$^{-1}$ and eV)
\begin{equation}\label{eq1}
k_f=0.512\sqrt{h\nu - E_B + V_0^*}.
\end{equation}
%Here, we assume free-electron like final states, whose final-state energies
%inside the material are 
%determined by 
Here the inner potential $V_0^*\approx 10$~eV is referenced to the Fermi energy and
the transferred photon momentum leads to a rigid shift of the free-electron final state sphere by the vector 
with absolute value $k_{h\nu}=2\pi\nu/c$ along the photon beam~\cite{Medjanik2017}.
The kinetic energy of the emitted photoelectrons is recorded by their time of flight and
the Fermi edge serves as reference for $E_B = 0$. The photon energy range used in these experiments is 560 to 660 eV (see Supplementary Material for details).
%The pattern observed on the detector represents the photoelectron intensity distribution as a function of the transversal momentum $k_{f,{||}}$. 

In our measurements, $I(E_B,k_x,k_y,k_z)$ is the intensity averaged between the two light polarizations. The intensity asymmetry, which contains the dichroism information, is calculated pixel-by-pixel as 
%\begin{equation}\label{asymmetry}
   $ A=(I_+-I_-)/(I_++I_-)$,
%\end{equation}
with $I_+$ and $I_-$ denoting the intensity measured at circular right and left polarization. This intensity asymmetry contains a well known dichroism component related to the measurement geometry and a component connected to the magnetic ordering. The %well understood 
geometry related component 
is the so-called circular dichroism in the angular distribution (CDAD)~\cite{Westphal1989,Daimon1995}, which is included  in the asymmetric component of
$A(k_x,k_y)$ with  respect to  the line $(k_x,k_y=0)$ coinciding with the $\Gamma - {\rm Z}$ direction.
CDAD is observed for a dissymmetric (handed) spatial arrangement of the quantization axis of initial state orbital momenta ($\bf{n}$), the
photon impact direction ($\bf{k}_{h\nu}$) and the photoelectron momentum ($\bf{k}_e$). Thus, CDAD from non-magnetic targets requires a handedness in the experimental geometry. It is therefore strictly antisymmetric with respect to the plane of photon 
incidence spanned by $\bf{k_{h\nu}}$ and the surface normal ([110] direction).
To isolate the CDAD in the experimental data we calculate the corresponding asymmetry as %$A_{\rm CDAD}$ by
%\begin{equation}\label{CDAD}
$A_{\rm CDAD}(k_x,k_y)=(A(k_x,k_y)-A(k_x,-k_y))/2$.    
%\end{equation}

In contrast to non-magnetic systems, magnetic systems can provide an additional asymmetry mechanism if the light polarization vector is parallel or antiparallel to the spin axis, which gives rise to MCD~\cite{Schneider1991}.
We can eliminate the contribution from the $A_{\rm CDAD}$ in the experimental asymmetry data, and hence extract the MCD contribution in
the remaining asymmetry, by calculating 
%\begin{equation}\label{MCD}
    $A_{\rm MCD}=(A(k_x,k_y)+A(k_x,-k_y))/2$.
%\end{equation}

We present next the key results of our studies in Fig.~\ref{Fig2JS}. We show the measured intensity $I(E_B,0,k_y)$ along the M-$\Gamma$-M line (see Fig. 1(c)), the intensity asymmetry $A(E_B,0,k_y)$, the MCD $A_{\rm MCD}(E_B,0,k_y)$, and the CDAD $A_{\rm CDAD}(E_B,0,k_y)$ in  Fig.~\ref{Fig2JS} (a)-(d), and their
corresponding  ab-initio based calculations for the specific experimental geometry in Fig.~\ref{Fig2JS} (e)-(h). The theoretical DFT+U calculations are based on
the one-step formulation of the photoemission process, using
the Korringa–Kohn–Rostoker ab-initio approach that represents the electronic structure of a system directly and efficiently in terms of its single-particle Green’s function
\cite{Ebert2011a,Braun2018}. The parameters used in the calculations correspond to those in Ref.~\onlinecite{Smejkal2022b}. 
We also note that  the MCD spectra is not a direct map of the ground state polarization, as shown in Ref.~\onlinecite{Scholz2013}, due to the final state effects. 
The calculations take into account the free electron like final
state at the corresponding $k_z$ (e.g. photon energy of 680eV) and the
matrix element of the induced transition. 
% since in general there is not a direct one-to-one correspondence between the spin-resolved band structure and the MCD spectra. 
As it is directly seen in Fig.~\ref{Fig2JS} (c) and (g), the experimentally measured and theoretically calculated MCD spectra show a very strong $\cal{T}$ symmetry breaking whose magnitude 
%can only arise from a 
is consistent with the exchange dominated mechanism as predicted by the theory of altermagnetism \cite{Smejkal2021a}. It is also important to contrast the MCD and CDAD data, which shows a dominance of the MCD contribution to the  intensity asymmetry, confirming its direct observation beyond any experimental artefact that may have originated from the CDAD signal. 

\begin{figure}
\includegraphics[width=\columnwidth]{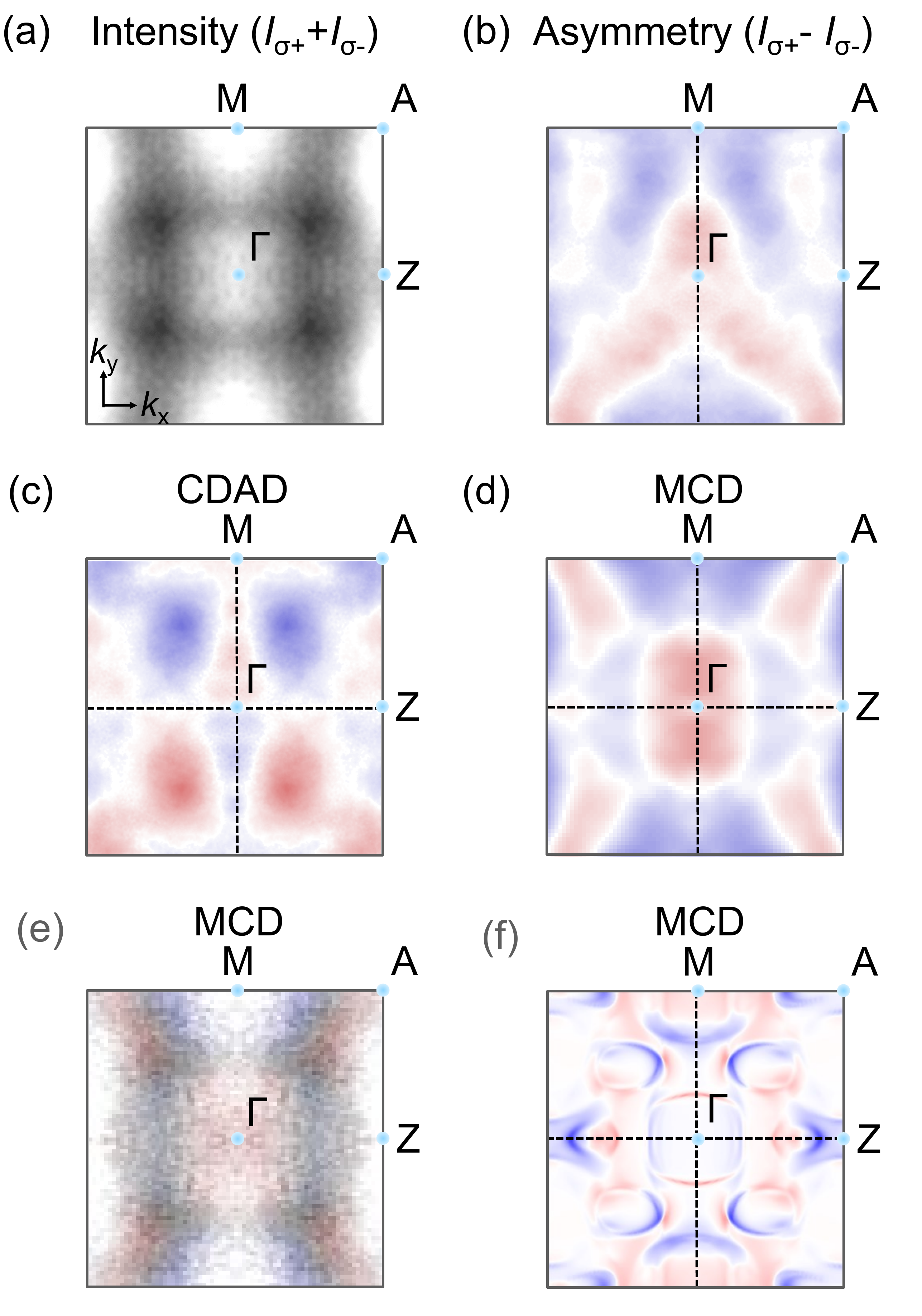}
\vspace{-0,0 cm}
\caption{\label{Fig3-JS} 
{(a) Constant energy map $I(E_F,k_x,k_y)$ measured 
at $h\nu= 380$~eV at 70~K on the $\Gamma$-M-A-Z plane.
The intensity has been averaged for circular left and right polarization. Here dark is high intensity.
(b) Asymmetry  of the constant energy map. 
(c) CDAD  map obtained from the asymmetry  in (b).
(d) MCD   map obtained from the asymmetry  in (b).
(e) MCD  map convoluted with the intensity map in (a).  
(f) Calculated MCD  map distribution (not convoluted with the intensity). The red-blue features in the 
$\Gamma$-M line are clear between (e) and (f). Features on the edges near Z are overestimated since they are
calculated on vanishingly small intensity features. }
}
\vspace{-0,4 cm}
\end{figure}

In Fig.~\ref{Fig3-JS} we present the measured intensity and asymmetry at the Fermi energy in the $\Gamma$-M-A-Z plane. 
The asymmetry  is shown in Fig.~\ref{Fig3-JS} (b) and the corresponding CDAD and MCD components in Fig.~\ref{Fig3-JS} (c) and (d). 
The convolution of the MCD with the average intensity plots is shown in Fig.~\ref{Fig3-JS} (e). It reveals the relevant parts of the MCD spectra, since some 
seemingly prominent features in Fig.~\ref{Fig3-JS}(d) %and the theory calculations, Fig.~\ref{Fig3-JS}(f), 
have very low intensity and hence are not directly reliably measured. 
For comparison, Fig.~\ref{Fig3-JS} (f) shows the theoretical MCD  (not convoluted with the intensity) at the Fermi energy.

We have confirmed the magnetic origin of 
the observed  $A_{\rm MCD}$ spectra by repeating the experiment after
rotating the sample around the surface normal by  180$^\circ$, which effectively rotates the magnetic order.
We present the experimental results and the corresponding theoretical calculations for both orientations in the Supplementary Material (see Figs.~\ref{Fig3-supp} and~\ref{Fig3-Theory}).
The distribution of $A_{\rm CDAD}$ is similar to the results for the non-rotated sample.
This can be expected because the ($k_x$,$k_z$) plane represents a crystal mirror plane.
In contrast, $A_{\rm MCD}$ reversed its sign as expected. The spectra do not match exactly since due to the experimental set-up limitations the 
area illuminated is not exactly the same, but  even at this semi-quantitive level the conclusion remains the same. 
This result also confirms that the geometry of the experiment points to a spin quantization
axis along $k_x$, corresponding to the c-axis of the RuO$_2$ crystal structure.

%%%%%%%%%%%%%%%%%%%%%%%%%%%%%%%%%%%%%%%%%%%%%%%%%%%%%%%%%%%%%%%%%%%%%%%%%%%%%%%%%
{\it Results with ultraviolet excitation} --
We have further confirmed the results by performing ultraviolet excitation experiments with a photon energy of 6.4~eV using  
an infrared fibre laser with quadrupled photon energy. 
%These experiments are favourable with respect to photon intensity. 
The results are restricted to a field of view limited to an area near the zone center [see Supplementary Material Fig.~\ref{Fig5}(b)].
We present the circular dichroism results obtained for 6.4~eV photon energy
 in the Supplementary Material  Fig.~\ref{Fig6}.
The asymmetries and decomposition in CDAD and MCD are calculated in the same way as for the soft X-ray results.
The ultraviolet excitation results are fully consistent with the X-ray results depicted above. 

\begin{figure*}[t]
\includegraphics[width=0.8\textwidth]{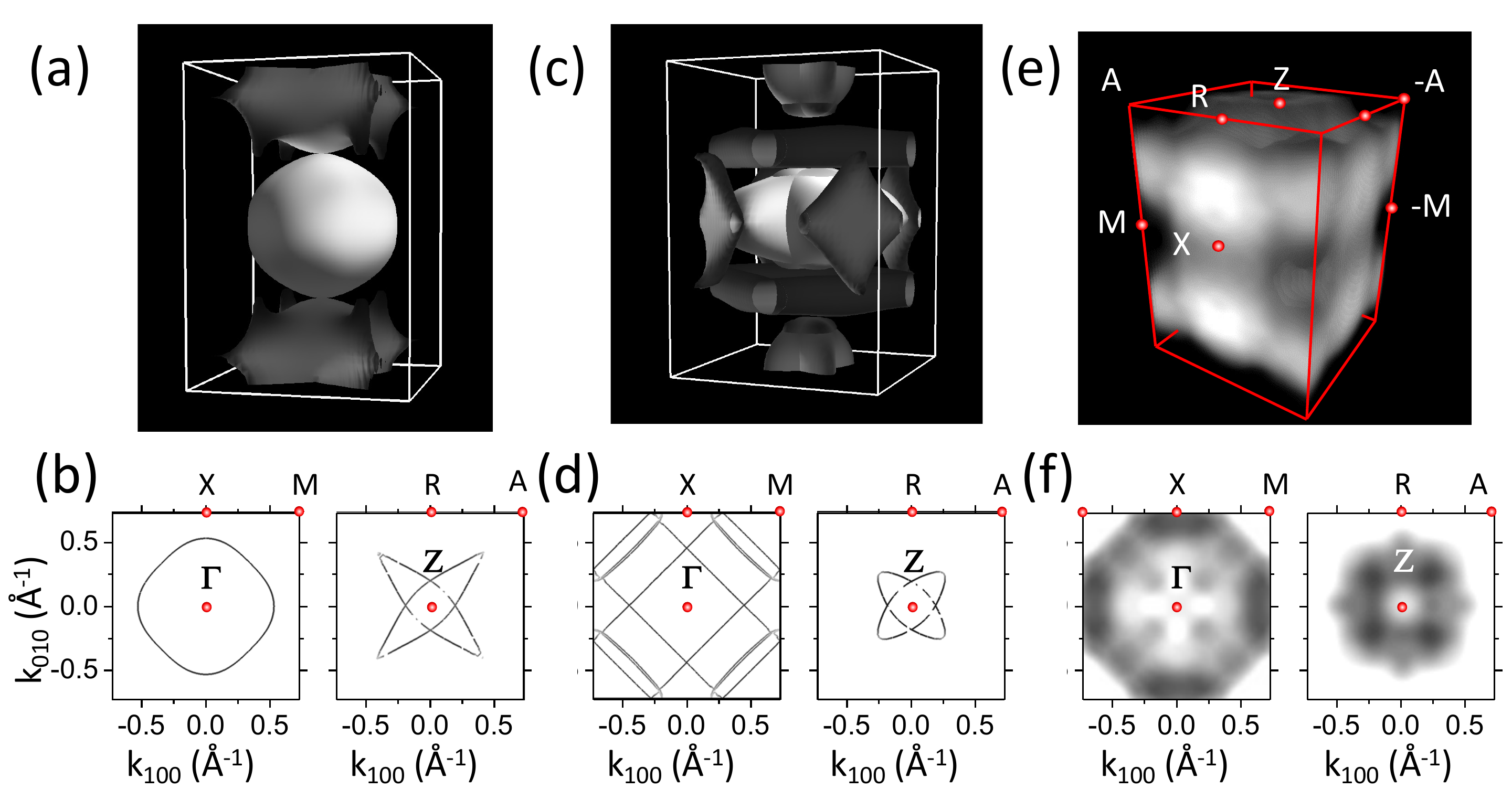}
\vspace{-0,5 cm}
\caption{\label{Fig4-JS} (a) Calculated Fermi surface for paramagnetic RuO$_2$. 
(b) Fermi surface cut for the plane perpendicular to the c-axis at the $\Gamma$-X-M plane and the Z-A-R plane for the paramagnetic phase.
(c) Calculated 
Fermi surface for the collinear magnetically ordered RuO$_2$. 
(d) Fermi surface cut for altermagnetic phase in the same planes as in (b).
(e) Fermi surface obtained experimentally
through a topographic mapping of the three-dimensional Brillouin zone (details in Supplementary Material).
(f) Photoelectron intensity at the Fermi energy
for planes as in (b).
}
\vspace{-0.5 cm}
\end{figure*}
In addition,  as a control test, we also performed the ultraviolet excitation photoemission experiment
with the c-axis orientated perpendicular to the incident light beam (see Supplementary Material  Fig.~\ref{Fig6}(f-j)).
In this case,  $A_{\rm MCD}$ [Fig.~\ref{Fig6}(j)] vanishes within error limits.
This observation indicates that the spin axis 
points perpendicular to the light polarization vector and hence parallel to the c-axis [001],
in  agreement with the results obtained with the soft X-ray excitation.

%%%%%%%%%%%%%%%%%%%%%%%%%%%%%%%%%%%%%%%%%%%%%%%%%%%%%%%%%%%%%%%%%%%%%%%%%

In order to verify  the altermagnetic phase of RuO$_2$, we can compare the experimental results with key theoretically predicted features for the paramagnetic vs. the altermagnetic phase. The theoretical calculations for the paramagnetic phase are shown in Fig.~\ref{Fig4-JS}(a), together with the Fermi surface cuts 
for the plane perpendicular to the c-axis [110] at the $\Gamma$-X-M plane and the Z-A-R plane in
Fig.~\ref{Fig4-JS}(b). We show the corresponding Fermi surface and cuts
for the altermagnetic phase in Fig.~\ref{Fig4-JS}(c),(d). The comparison to experiment is obtained by 
 a tomographic mapping of the three-dimensional Brillouin zone 
[Fig.~\ref{Fig0}(c)], obtained
by varying the photon energy in the range of 560 - 660~eV. According to Eq.~\ref{eq1}
this variation results in $k_z=k_{[110]}$ values ranging from 
$5G_{[110]}$ to
$5.5G_{[110]}$, i.e., from the center to the rim of a Brillouin zone 
[see Supplementary Material Fig.~\ref{Fig4-v2}(b)]. Here $G_{[110]}$ is the magnitude of the [110] reciprocal lattice vector.
Exploiting the translational symmetry in momentum space, the intensity distributions $I(E_B,k_x,k_y,k_z)$ map a complete Brillouin zone. 
The Fermi energy intensity distribution shown along the same cuts and the overall shape of the reconstructed Fermi surface in Fig.~\ref{Fig4-JS} (e),(f)
match directly with the theoretical calculations of the collinear compensated altermagnetic phase in Fig.~\ref{Fig4-JS} (c),(d).

{\it Conclusion} --
%Using photon excitation in the soft X-ray and ultraviolet regime,
%we have investigated circular dichroism in the angular distribution of photoelectrons.
%Part of the dichroism can be explained by the 
%classical CDAD. CDAD represents a pure non-relativistic effect and 
%does not depend on the spin orientation.
%It implies a strictly antisymmetric behavior with respect to the plane spanned
%by the surface normal and the incident photon vector coinciding with a crystal mirror plane.
%For both photon energy ranges we observe in addition a non-vanishing asymmetry component being
%symmetric with respect to the plane of photon incidence.
%This magnetic circular dichroism, MCD, requires a spontaneous broken
%symmetry of the sample like in the case of ferromagnetism.
%The experimental geometries imply 
%that the antiferromagnetic moments are oriented parallel to the c-axis of the sample.
%Because the crystal structure itself does not provide any chirality, we attribute the MCD to the
%itinerant antiferromagnetic band structure of altermagnetic RuO$_2$.
We have experimentally established the key signature of the recently predicted \cite{Smejkal2021a,Smejkal2022a} altermagnetic phase  by directly detecting $\cal{T}$-symmetry breaking in the band structure of the collinear compensated magnet RuO$_2$. Supported by ab initio calculations, our experimental results underpin on the microscopic electronic-structure level the recently reported unconventional macroscopic responses, namely the anomalous Hall and spin-polarized currents accompanied by vanishing magnetization \cite{Feng2022,Betancourt2021,Bose2022,Bai2022,Karube2022}, in this workhorse altermagnetic material.  In general, our results microscopically establish the grounds for the exploration and exploitation of envisaged \cite{Smejkal2021a,Smejkal2022a} phenomena and functionalities based on the altermagnetic $\cal{T}$-symmetry breaking that are beyond the reach of the conventional magnetic phases in fields ranging from spintronics, ultrafast magnetism, magneto-electrics, and magnonics, to topological matter and superconductivity.

%We experimentally established the key signature of the recently predicted altermagnetic phase by microscopically vizualizing $\cal{T}$ symmetry breaking in the band structure of the collinear compensated magnet  RuO$_2$.
%
%We have directly measured the $\cal{T}$ symmetry breaking in the band structure of the altermagnetic RuO$_2$
%through the MCD effect in the angular distribution
%of photoelectrons.
%The measured asymmetry texture on the three-dimensional Fermi surface coincides
%with theoretical predictions and calculations for this metallic altermagnet. 

%Furthermore, we envision the application of MCD to reveal magnetization structures and domain walls in these itinerant collinear altermagnets using 
%photoemission electron microscopy by locating the contrast aperture in an asymmetric position,
%{\it i.e.} dark field-imaging.

Sincere thanks are due to M. Kallmayer and A. Oelsner (Surface Concept GmbH) 
and T. Grunske, T. Kauerhof, and K. von Volkmann (APE GmbH) for continuous support.
This work was funded by the Deutsche Forschungsgemeinschaft (DFG) Grant No. TRR 173 268565370 (projects A02, A03, A05, and B02), by the BMBF  (projects 05K22UM1 and 05K19UM2), by JGU TopDyn initiative, by the EU FET Open RIA Grant no. 766566, and by the Grant Agency of the Czech Republic grant no. 19-28375X.
J.M. and S.W.DS would like to thank CEDAMNF project financed by the Ministry of Education, Youth and Sports of Czech Republic, Project No. CZ.02.1.01/0.0/0.0/15$\_$003/0000358.
Q.N. acknowledges support by the Q-FARM Bloch Fellowship and U.S. DOE (DE-AC02–76SF00515).

%\bibliography{RuO2bib,MendeleyBib}

\begin{widetext}

\section{Supplementary Material}

\subsection{Sample preparation and characterization}

Epitaxial RuO$_2$(110) films with a thickness of 34~nm were grown by pulsed laser deposition on
TiO$_2$(110) substrates that were heated during deposition to 400 $^o$C. 
The oxygen pressure was 0.02~mbar and typical growth rates were 1.9~nm/min with the KrF excimer laser running at 10~Hz and 150~mJ pulse energy. Samples 
 were structurally characterized by X-ray diffraction, X-ray reflectometry and in-situ reflective high energy electron diffraction (RHEED). From  X-ray reflectometry we find that 34~nm samples have a surface roughness smaller than 0.3~nm, while the width of the out-of plane scattering peak in 2$\Theta$ indicates a somewhat reduced coherent scattering volume of 22~nm. The existence of the RHEED scattering pattern shows that crystallinity persists up to the surface. The full width at half maximum of the rocking curve is 1.0$^o$ indicating a good alignment of the surface normal. $\Phi$ scans reveal the epitaxial growth and show a two fold symmetry of (200) peaks aligned with that from the substrate. The width of the peaks in $\Phi$ is 2$^o$. 
 From the determined lattice constants we conclude on a compressive strain of -1.4\% along $c$-axis and a tensile strain of 0.4\% along $a$-axis. 
Samples were transported from the deposition chamber to the photoemission experiment  using an ultra-high vacuum suitcase.

\setcounter{figure}{0}
\makeatletter 
\renewcommand{\thefigure}{SM-\@arabic\c@figure}
\makeatother

\begin{figure}
\includegraphics[width=0.9\columnwidth]{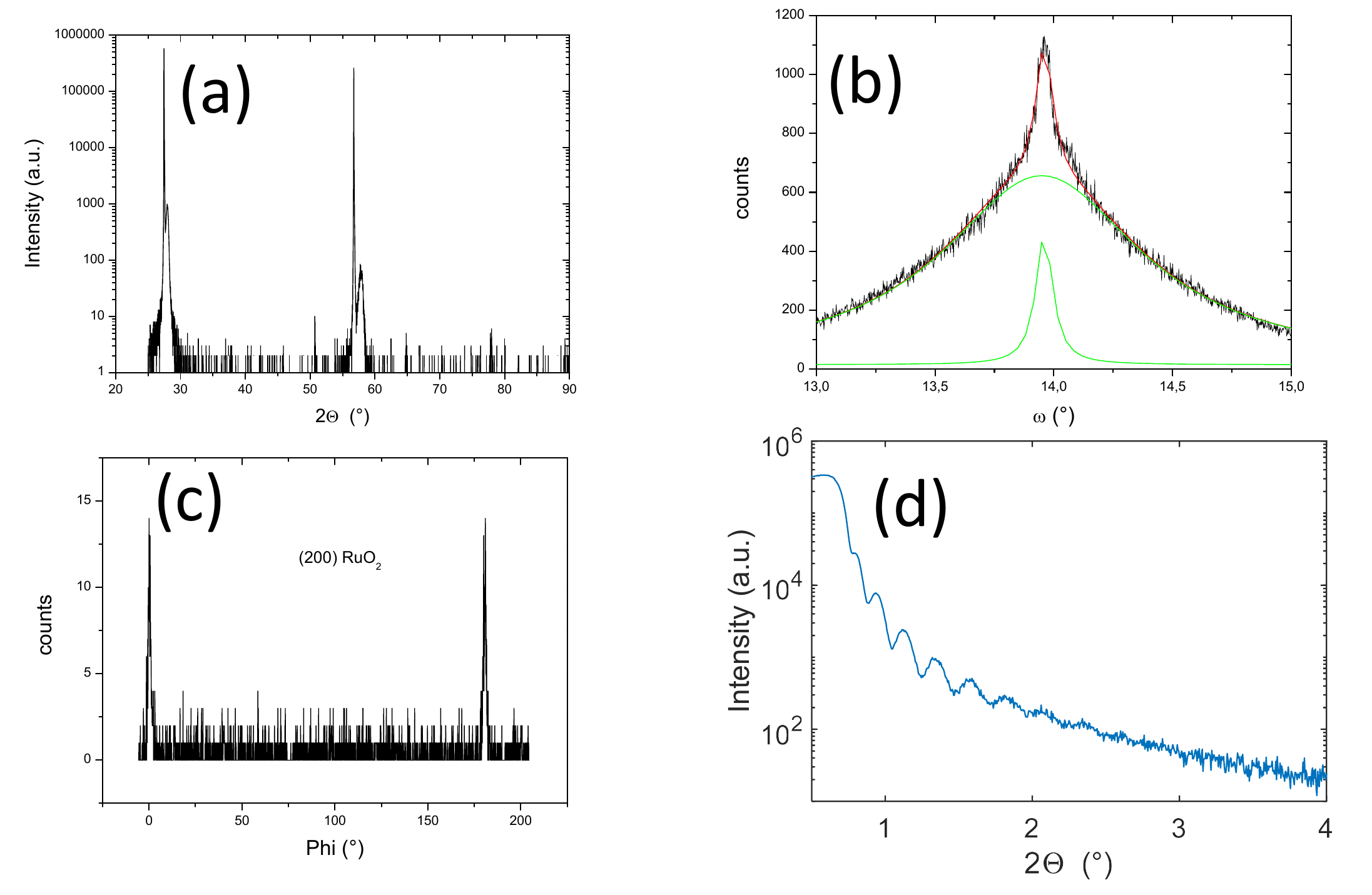}
\vspace{-0,6 cm}
\caption{\label{FigXRD} 
(a) $\Theta/2\Theta$ scan of RuO$_2$ [110] oriented film. (b) Rocking curve of RuO$_2$ [110] oriented film. Black line is measured data; red line sum of the two fitted Lorentz lines (green). (c) $\phi$ (Phi) scan of non specular [200] RuO$_2$ film.  (d) X-ray reflectivity of RuO$_2$ film.  
}
%\vspace{-0,6 cm}
\end{figure}
We examined the crystal structure of the samples using a four circle Bruker D8 Discover X-ray diffractometer. The instrument was operated with a copper anode, a Goebel mirror, and a Germanium 2-bounce monochromator to eliminate Cu K$_{\alpha2}$ and Cu K$_\beta$ radiation. Fig.~\ref{FigXRD}(a) displays a $\Theta/2\Theta$ scan of the sample investigated at the synchrotron. The figure reveals pronounced peaks corresponding to the [110] and [220] orientations of the substrate. Adjacent to them, we observe film peaks corresponding to [110] and [220], indicating slightly smaller lattice plane distances for the films compared to the substrate. No other peaks are visible in the X-ray diffraction diagram. The small yet distinct peak at 50.7$^\circ$ results from Cu K$_\beta$ radiation that was not perfectly filtered and diffracted at the [220] substrate peak.

The analysis of the tilt of [110] film lattice planes relative to the surface normal was conducted through the performance of rocking curves ($\omega$-scans), as depicted in Fig.~\ref{FigXRD}(b). The observed peak exhibited two contributions and was fitted using a summation of two Lorentzian lines. The exceptionally sharp peak is presumed to originate from the nearby substrate peak. The second peak, characterized by a width of 1.0$^\circ$, provides information regarding the tilt of the out-of-plane directions of the crystallites. The in-plane alignment of the crystallites was confirmed through $\phi$ (Phi) scans of non-specular reflections, such as the [101] and [200] film peaks. In Fig.~\ref{FigXRD}(c), only two reflections with a 180$^\circ$ difference are observed, specifically the [200] and [020] film peaks. The $\phi$ angles are aligned with their corresponding substrate reflections.

Low-angle X-ray reflectivity experiments provide valuable information regarding the total film thickness and film roughness. By analyzing the distance between oscillations in Fig.~\ref{FigXRD}(d), we calculated a film thickness of 33.7 nm and a total roughness of 1.5 nm, which includes contributions from both substrate and surface roughness.

\subsection{Details of  soft X-ray excitation}

As we have described in the main text,
using soft X-ray excitation measures the photoelectron intensity distribution $I(E_B,k_x,k_y,k_z)$.
Considering the energy conservation and the role of reciprocal
lattice vectors in order to obey momentum conservation,
for a given photon energy $h\nu$ and binding energy $E_B=E-E_F$,
the final photoelectron states are located on a 
spherical shell with radius (for units \AA$^{-1}$ and eV)
\begin{equation}\label{eq2}
k_f=0.512\sqrt{h\nu - E_B + V_0^*}.
\end{equation}
Here, we assume free-electron like final states, whose final-state energies
inside the material are 
determined by the inner potential $V_0^*\approx 10$~eV referenced to the Fermi energy.
The transferred photon momentum leads to a rigid shift of the final state sphere by the vector 
with absolute value $k_{h\nu}=2\pi\nu/c$ along the photon beam~\cite{Medjanik2017}.

The kinetic energy of the emitted photoelectrons is recorded by their time of flight. 
The Fermi edge serves as reference for $E_B = 0$. 
The pattern observed on the detector represents the photoelectron intensity distribution as a function of the transversal momentum $k_{f,{||}}$. 

We next discuss the photoelectron intensity distributions averaged for circular left and right polarization.
Fig.~\ref{Fig1}(a,b) shows results recorded at a photon energy of 380~eV corresponding to final states in the 5th repeated BZ along
the direction perpendicular to the surface ($k_z$). 
The results are shown as raw data divided by the detector response function.
The constant energy map [Fig.~\ref{Fig1}(a)] at the Fermi level reveals the two-fold symmetry of the RuO$_2$(110) surface.
The cut along the high symmetry direction $\Gamma - Z$ indicates the energy dispersion of the valence band $E_B(k_x)$ [Fig.~\ref{Fig1}(b)].
We observe a homogeneous background intensity and
the maximum photoelectron intensity of direct transitions amounts to 20\% of the background intensity, originating from quasi-inelastic scattering.
The electron bands appear broadened both on the energy and momentum scale, which cannot be explained by the finite energy and momentum resolution of the instrument.
% It can also not be explained by structural imperfectness of the samples because X-ray diffraction revealed high structural coherence.
% GJ: I commented previous sentence since I now think there is some strain relaxation in the sample
% as Scherrer formula gives a smaller grain size for 220 than for 110 and due to shape of rocking curves
% I will make some simulations of XRD data but this might take time
Instead, we attribute the broadening to the electron correlation, which is expected to be large in oxides~\cite{Ahn2019}.

\begin{figure}
\includegraphics[width=0.5\columnwidth]{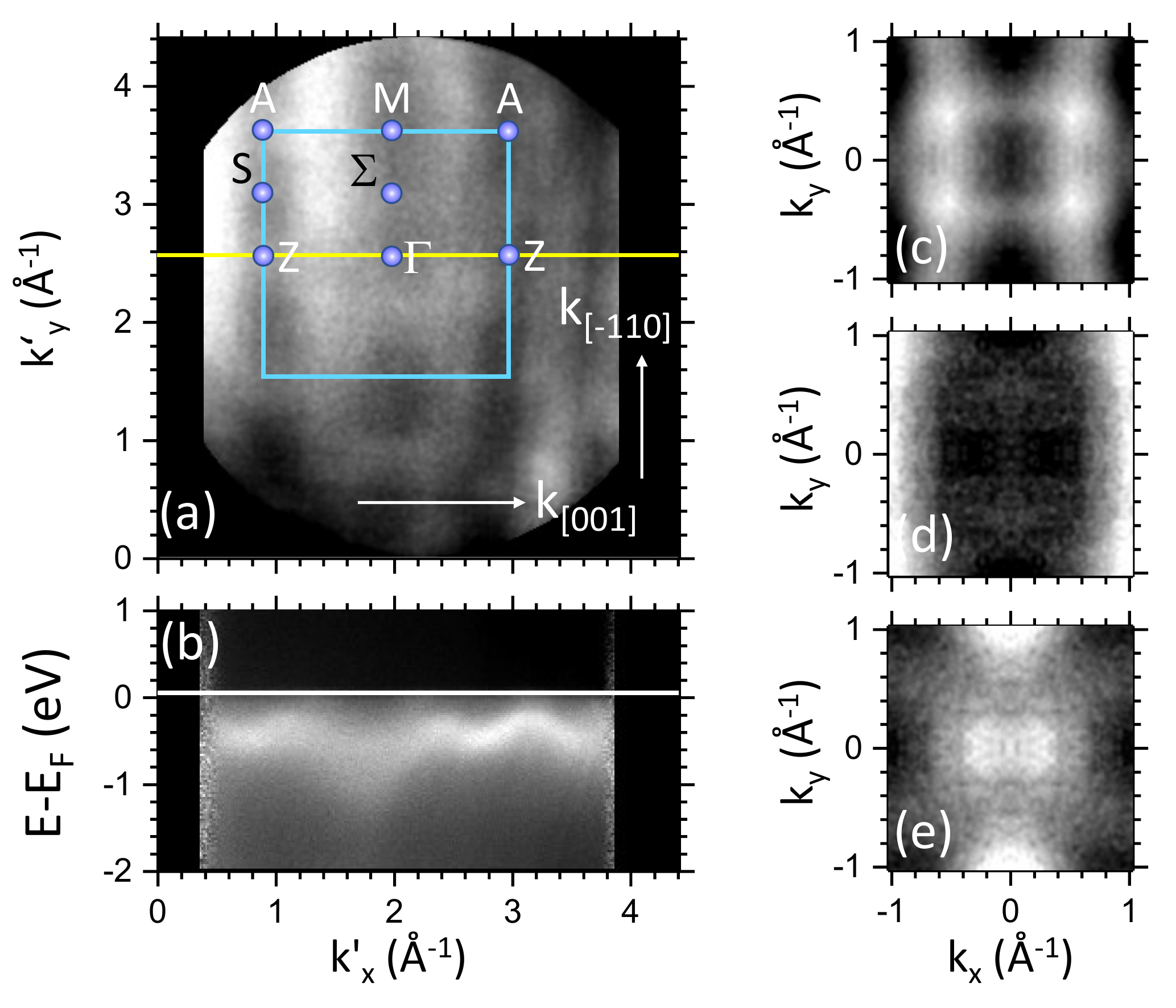}
\vspace{-0,6 cm}
\caption{\label{Fig1} 
(a) Constant energy map $I(E_F,k_x,k_y)$ measured 
at $h\nu= 380$~eV at 70~K. 
The intensity has been averaged for circular left and right polarization. 
The photon beam impinges from the right parallel to $k_x$. The blue square
indicates the Brillouin zone boundary.
(b) Intensity map $I(E_B,k_x)$ revealing the energy dispersion of bands along the profile
indicated by the horizontal yellow line in (a).
(c-e) Symmetrized constant energy maps covering one Brillouin zone at 
binding energies (c) $E_B=0$, (d) 0.5, and (e) 1.0~eV 
at $h\nu= 660$~eV.
}
%\vspace{-0,6 cm}
\end{figure}
The symmetrized constant energy maps centered at $\Gamma$ in the range of 
$E_B=0 - 1$~eV [Fig.~\ref{Fig1}(c-e)] have been acquired for 
$h\nu=660$~eV and cover one Brillouin zone. 
The maps agree with previously published results~\cite{Jovic2018} on RuO$_2$ single crystals obtained with a photon energy of 131~eV, except for the localized high intensity at the crossing points of horizontal and vertical lines, which may stem from surface states.
Hence, we attribute the observed photoelectron intensity distribution to bulk states of RuO$_2$.

\begin{figure*}
\includegraphics[width=0.9\columnwidth]{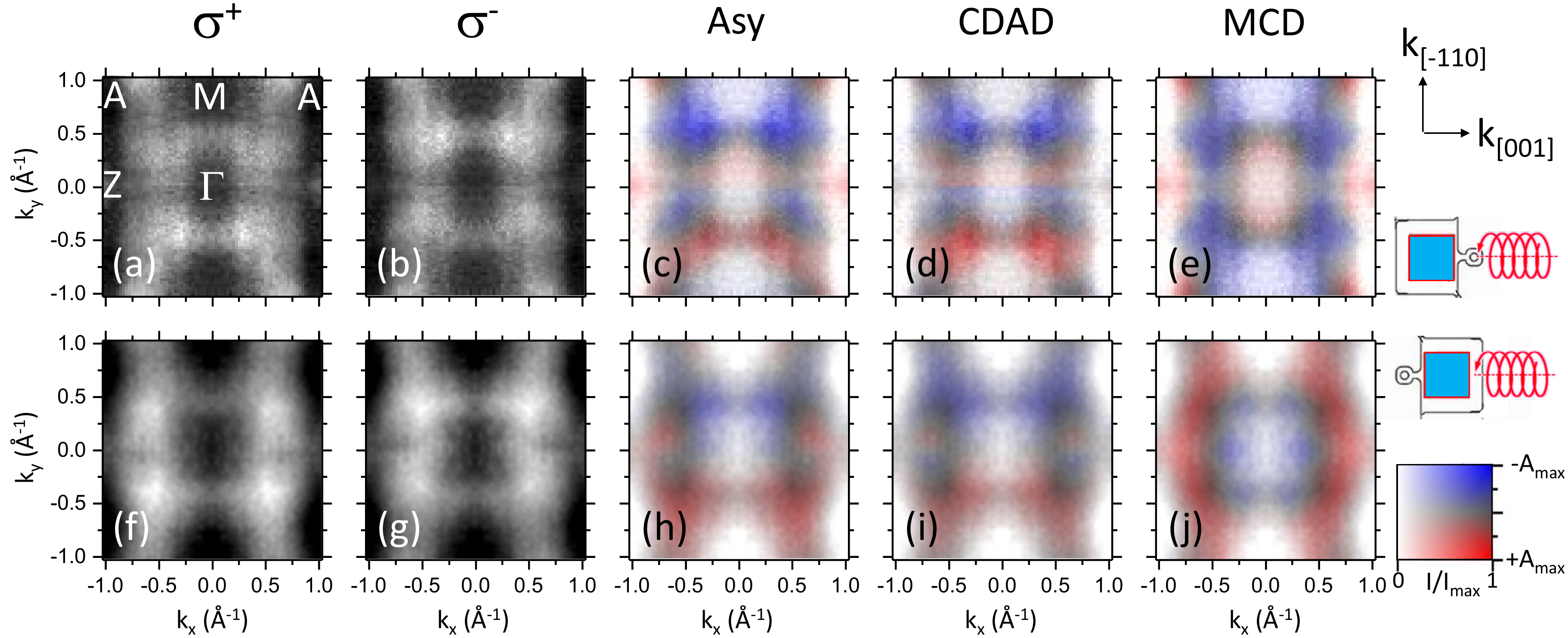}
\vspace{-0,0 cm}
\caption{\label{Fig3-supp} 
(a,b) Photoelectron intensity maps at the Fermi energy obtained with
right ($\sigma^+$) and left ($\sigma^-$) circularly polarized light.
(c) Asymmetry and averaged intensity depicted in a combined color scale.
(d) CDAD and intensity in the same color scale.
(e) MCD and intensity depicted in a combined color scale.
(f-j) Similar data for the sample being rotated by 180 degrees
as indicated by the sketches on the right.
}
%\vspace{+0,6 cm}
\end{figure*}
To further verify the results of the circular dichroism observations, we show the  intensity maps at the Fermi level for circular left and right polarization [Fig.~\ref{Fig3-supp}(a,b)]. Here, as compared to the main text, the asymmetry, CDAD, and MCD are presented convoluted with the averaged intensity maps [Fig.~\ref{Fig3-supp}(c)-(e) and (h)-(f)]. As with the results presented in Fig.~\ref{Fig2JS}, the asymmetry  is decomposed in the CDAD and MCD components. 
The result shown in Fig.~\ref{Fig3-supp}(d) reveals predominantly negative values for positive $k_y$ except for a reversed sign close to $k_y=0$.
The maximum experimental values amount to $\pm$5.4\%. 
After subtracting the homogeneous background intensity, the
CDAD for direct transitions is $\pm$27\% and its size is thus in the range of previously reported values~\cite{Schoenhense1990}.
For non-magnetic systems $A_{\rm CDAD}$ is the only possible circular dichroism mechanism except for crystal structures with natural chirality, where effects are  extremely small.
On the other hand, CDAD is ubiquitous in photoemission because it is a pure orbital effect and does not require spin-orbit interaction and heavy elements.

The MCD data show a negative asymmetry for the vertical stripes parallel to $\Gamma$-M and
positive values for the horizontal strips parallel to $\Gamma$-Z.
Along the M-$\Gamma$-M path in Fig. 6(e), $A_{\rm MCD}$ is negative near the M-points and positive near the $\Gamma$ point, in good agreement with the results shown in Fig. 2(c) for the Fermi level.
%The maximum values of $A_{\rm MCD}$ are 
%%$\pm$3\% and after subtracting the background intensity 
%$\pm$15\%.
%$A_{\rm MCD}$ is thus remarkably large compared to values of a few percent reported for $3d$ metallic ferromagnets~\cite{Schneider1991}.

To confirm that the observed  $A_{\rm MCD}$ is connected to
the magnetic order of the sample, we rotated the sample around the surface normal by 
180 degrees and repeated the photoemission experiment, albeit not in the same exact spot given the experimental limitations of the set-up.
Corresponding results are shown in Figs.~\ref{Fig3-supp}(f-j).
The distribution of $A_{\rm CDAD}$ is similar to the results for the non-rotated sample.
This can be expected because the ($k_x$,$k_z$) plane represents a crystal mirror plane.
In contrast, $A_{\rm MCD}$ has reversed its sign. The reversal of the sign of the MCD data further confirms that the 
observations and those in Fig.~\ref{Fig2JS} are a directly observation of the
broken time-reversal symmetry of the altermagnetic sample.

In Fig.~\ref{Fig3-Theory} we present the theory calculations corresponding to the experiment and the spin-resolved bands presented in~\ref{Fig3-Theory} (e) and (j). Here the color plots are not convoluted with the intensity for easier visualization. Fig.~\ref{Fig3-Theory} (a)-(e) corresponds to the experimental set-up of \ref{Fig3-supp} (a)-(e) and Fig.~\ref{Fig3-Theory} (f)-(j) corresponds to the experimental set-up of \ref{Fig3-supp} (f)-(j). 

\begin{figure*}
\includegraphics[width=0.9\textwidth]{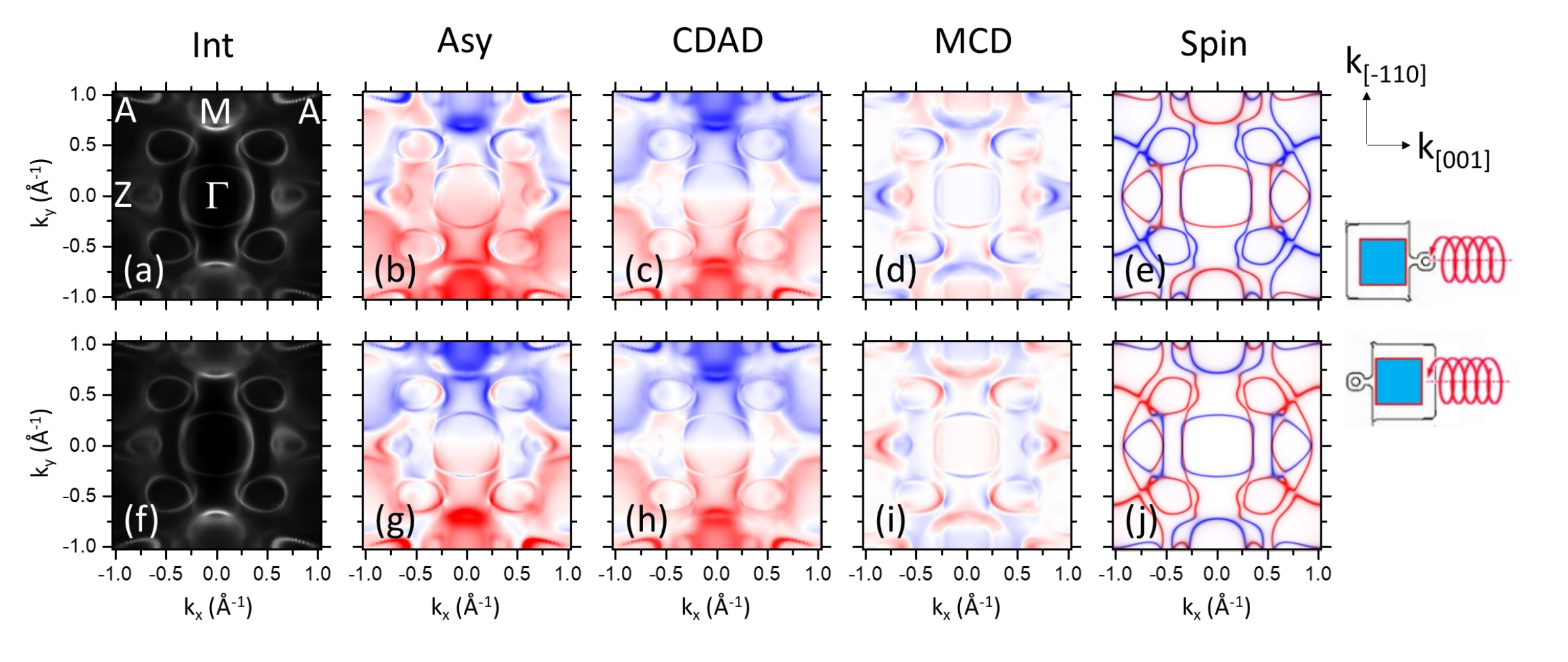}
\vspace{-0,0 cm}
\caption{\label{Fig3-Theory} 
(a) Theory calculation of photoelectron intensity maps at the Fermi energy.
(b) Asymmetry  depicted in a color scale.
(c) CDAD in the same color scale.
(d) MCD  in the same color scale.
(e)  Spin-resolved bands.
(f-j) Similar data for the sample being rotated by 180 degrees
as indicated by the sketches on the right.
}
\vspace{+0,6 cm}
\end{figure*}

We present the scan of the perpendicular momentum by photon energy in Fig.~\ref{Fig4-v2}. This topographic mapping is made by varying the photon energy in the range of 560 - 660~eV. According to Eq.~\ref{eq2}
this variation results in $k_z=k_{[110]}$ values ranging from 
$5G_{[110]}$ to
$5.5G_{[110]}$, i.e. from the center to the rim of a Brillouin zone 
[see Fig.~\ref{Fig4-v2}(b)].
Similarly, corresponding asymmetry maps were determined.
Exploiting the translational symmetry in momentum space, the intensity distributions $I(E_B,k_x,k_y,k_z)$ map a complete Brillouin zone. This topographic map and cuts are the ones used in Fig.~\ref{Fig4-JS}.
\begin{figure*}
\includegraphics[width=0.8\textwidth]{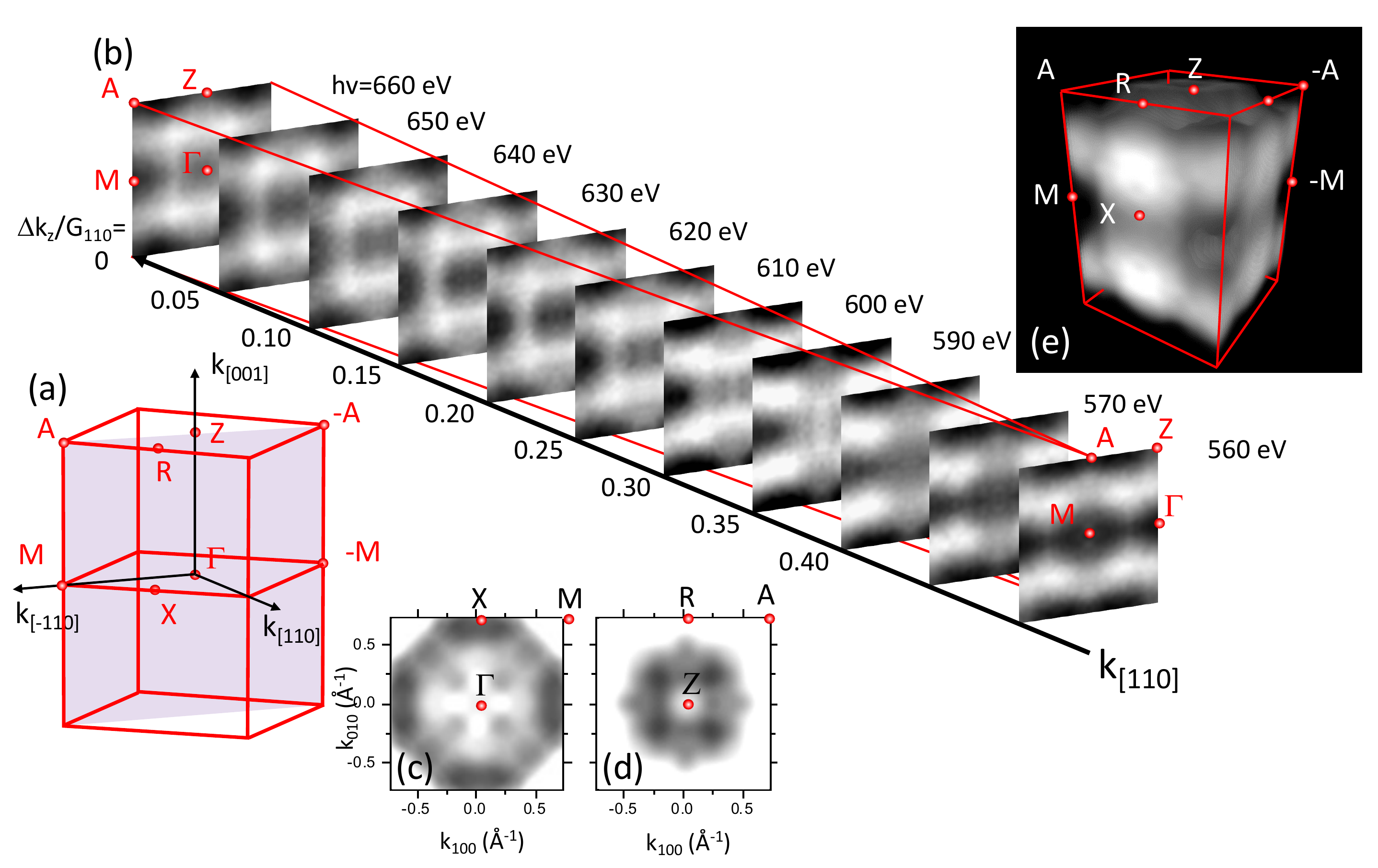}
\vspace{-0,0 cm}
\caption{\label{Fig4-v2} 
(a) Sketch of the Brillouin zone of the RuO$_2$ rutile structure in momentum space.
(b) Series of photoelectron intensity maps at the Fermi level for indicated photon energies. 
(c-d) Photoelectron intensity at the Fermi surface
for planes perpendicular to the c-axis [110] at $k_z=0$ and 0.5$G_{110}$.
(e) Three-dimensional Fermi surface.
}
\vspace{-0,0 cm}
\end{figure*}

\subsection{Results with ultraviolet excitation}
As compared to the previous results, photoelectron excitation with a photon energy of 6.4~eV using  
an infrared fibre laser with quadrupled photon energy is favourable with respect 
to photon intensity. Yet, the small photon energy limits the detectable parallel momentum
to $k_{||}<0.6$~\AA$^{-1}$ [see Fig.~\ref{Fig5}(a)].
Assuming an excitation into free-electron like final states~\cite{Stejskal2021}, 
Eq.~\ref{eq2}
results in a cut close to the center of the second repeated Brillouin zone
($\Gamma$) in momentum space  [see Fig.~\ref{Fig5}(b)].
At increased binding energy, we still observe the two pairs of horizontal and vertical
bands, which we identify with the band features seen with soft X-ray excitation.
The dispersion maps [Figs.~\ref{Fig5}(c,d)] reveal a maximum binding energy of the parabolic band at
$\Gamma$ of 0.6~eV, in reasonable agreement with results obtained with soft X-ray excitation.
When comparing band dispersions at different photon energies, 
the different cuts in the three-dimensional momentum-space and a possible variation of the effective mass must be considered.

\begin{figure}
\includegraphics[width=0.5\columnwidth]{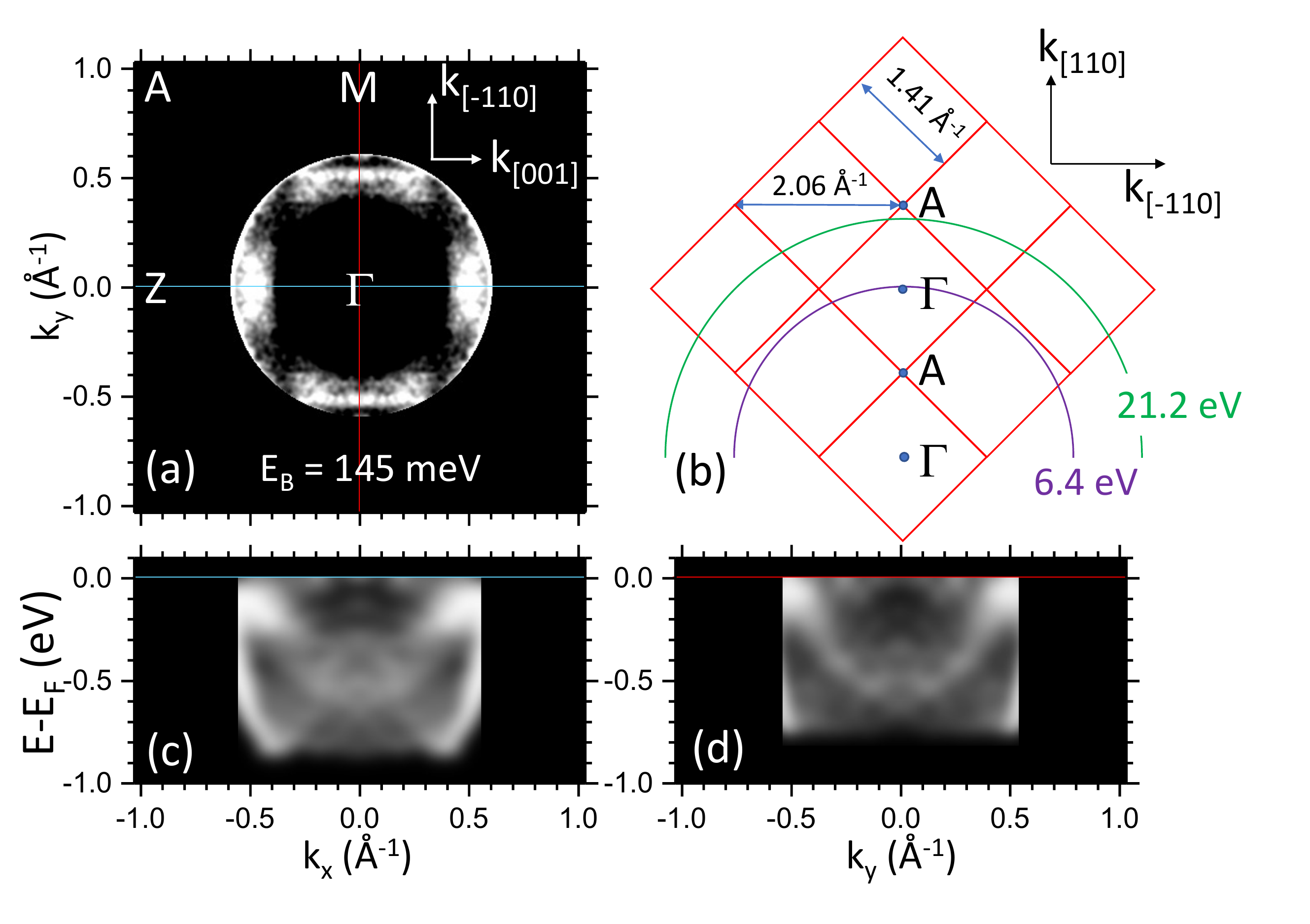}
\vspace{-0,0 cm}
\caption{\label{Fig5} 
(a) Polarization-averaged photoelectron intensity map at a binding energy of 145~meV for excitation with a photon energy of 6.4~eV. The photon beam impinges from the right along the in-plane [001] direction
of the RuO$_2$(110) film.
(b) Sketch of the repeated Brillouin zone scheme in a plane perpendicular to the [001] axis.
Green and violet half circles indicate the free-electron final state momenta for
photon energies 21.2~eV and 6.4~eV, respectively.
(c) Photoelectron intensity map in the $E_B$ vs. $k_x$ plane revealing the dispersion
close to the $\Gamma$-Z direction.
(d) Same, but along the $\Gamma$-M ($k_y$) direction, corresponding to
the violet half-circle in (b).
}
\vspace{-0,0 cm}
\end{figure}

\begin{figure*}
\includegraphics[width=\textwidth]{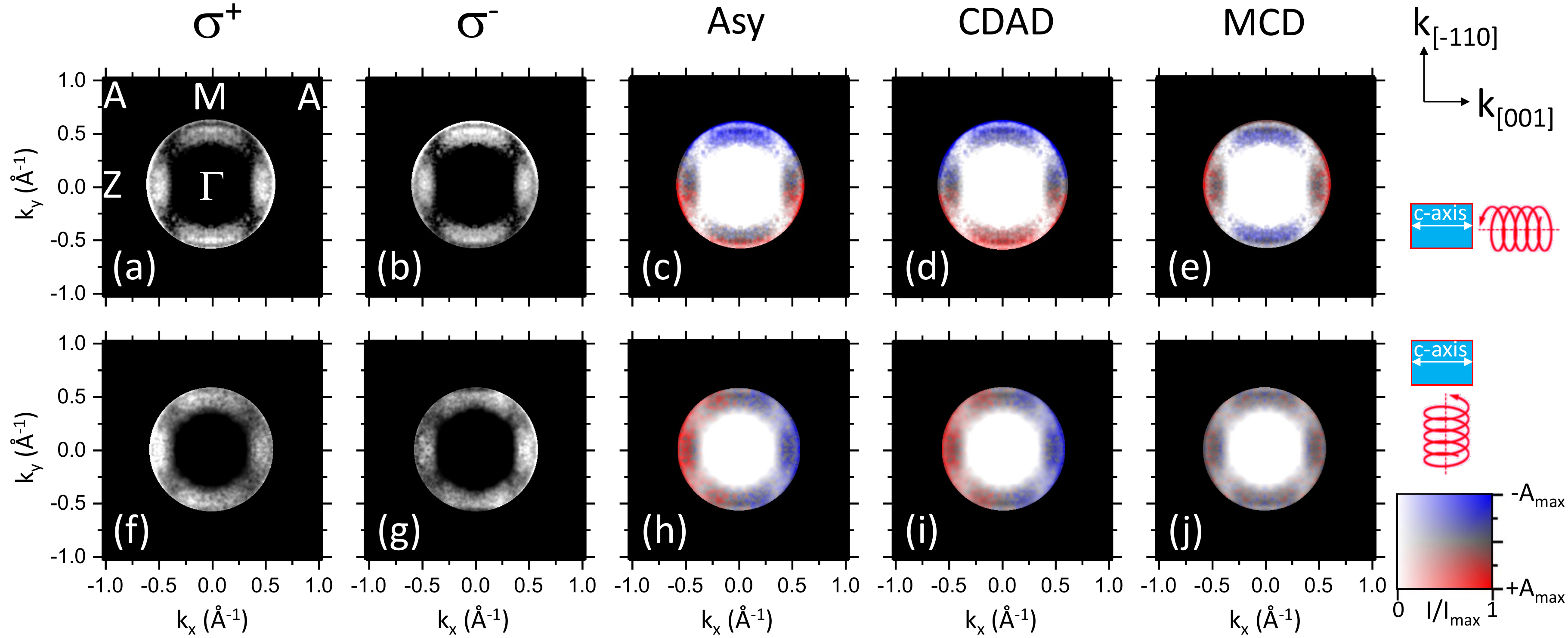}
\vspace{-0,0 cm}
\caption{\label{Fig6} 
(a,b) Photoelectron intensity maps at a binding energy of 182~meV obtained with
6.4~eV photon energy and circularly right ($\sigma^+$) and left ($\sigma^-$) polarized light.
(c) Asymmetry and averaged intensity plotted in a combined color scale.
(d) CDAD and intensity depicted in the same color scale.
(e) Same, but for the corresponding MCD and intensity.
(f-j) Similar data for 90 degrees rotated sample as indicated by the sketches on the right.
}
\vspace{+0,0 cm}
\end{figure*}
The circular dichroism results obtained for 6.4~eV photon energy
are shown in Fig.~\ref{Fig6}.
The asymmetries are calculated in the same way as for the soft X-ray results.
In this case the azimuthal orientation could not be varied in-situ and hence we compare to similarly prepared samples 
with the c-axis oriented parallel and perpendicular to the incident light beam.
For the case of parallel orientation [Fig.~\ref{Fig6}(a-e)]
$A(k_x,k_y)$ shows the expected antisymmetric behavior with respect
to the $\Gamma - {\rm Z}$ symmetry axis, originating from
 non-relativistic CDAD~\cite{Westphal1989,Daimon1995}.
By calculating
$2A_{\rm CDAD}(k_x,k_y)=A(k_x,k_y)-A(k_x,-k_y)$ we separate the CDAD asymmetry as shown in
Fig.~\ref{Fig3-supp}(c-d and h-i).
The CDAD shows the same antisymmetry as in the soft X-ray case
with the characteristic feature of a sign change at the line $k_y=0$.
This line corresponds to the coplanar geometry of crystal mirror plane and photon incidence plane.
The maximum experimental values amount to 
%$\pm$0.4\%. 
%After subtracting the homogeneous background intensity, the
%CDAD for direct transitions is 
$\pm$26\%.

The calculation of  $2A_{\rm MCD}=A(k_x,k_y)+A(k_x,-k_y)$ results in the magnetic contribution to
the circular dichroism [Fig.~\ref{Fig6}(e)].
We observe a positive asymmetry for the vertical stripes parallel to $\Gamma$-M and
negative values for the horizontal strips parallel to $\Gamma$-Z.
The maximum values of $A_{\rm MCD}$ are 
%$\pm$0.2\% and
%after subtracting the background intensity 
$\pm$13\%.
$A_{\rm MCD}$ observed with 6.4~eV excitation thus confirms
the non-vanishing magnetic circular dichroism observed for soft X-ray excitation.

A similar photoemission experiment
with the c-axis orientated perpendicular to the incident light beam serves as a control test
[Fig.~\ref{Fig6}(f-j)].
In this case, $A_{\rm CDAD}$ shows a left-right asymmetry with respect to
the $\Gamma$-M axis [Fig.~\ref{Fig6}(i)]. 
$A_{\rm MCD}$ [Fig.~\ref{Fig6}(j)] vanishes within error limits.
This observation indicates that the magnetization axis 
points perpendicular to the light polarization vector and hence parallel to the c-axis [001],
in perfect agreement with the results obtained with soft X-ray excitation.

\subsection{Ab initio  calculations}
We have calculated the ground state electronic structure of RuO$_2$ in P4$_2$/mnm (Space group:136) symmetry using the optimized lattice parameter (a = b = 4.5331 \AA, c = 3.1241 \AA) \cite{Smejkal2020}. These calculations were carried out using spin-polarized relativistic Korringa-Kohn-Rostoker (SPRKKR) Green's function method in the atomic sphere approximation (ASA), within the rotationally invariant GGA+U scheme as implemented in the SPRKKR formalism \cite{Ebert2011a,SuppMat-theory2}. The screened on-site Coulomb interaction U and exchange interaction J of Ru are set to 2.00 eV and 0.70 eV, respectively. The angular momentum expansion up to $l_{\rm max} = 4$ has been used for each atom on a 22 x 22 x 32 k-point grid. The energy convergence criterion has been set to 10-5 Ry. Lloyd’s formula has been employed for accurate determination of the Fermi level \cite{SuppMat-theory}.
The ab initio photoemission calculations of RuO2(110) were performed within the one-step model of photoemission in the spin-density-matrix formulation as implemented in the SPRKKR package \cite{Braun2018}, taking into account all geometry and light-induced effects of the photoemission process for the actual experiment including photoelectron angular distribution, matrix elements and final states constructed as the time-reversed LEED states. The final-state damping was described via constant $V_i = 0.1$ eV set to simulate finite inelastic mean free path. The electronic structure and Fermi surface calculations in Figs.~\ref{Fig0} and \ref{Fig4-JS} were calculated using FLAPW ELK code \cite{elk}. The nonmagnetic and magnetic calculations were performed without spin-orbit coupling.

\end{widetext}

\end{document}